\documentclass{article}
\usepackage{amssymb}
\usepackage{amsfonts}

\input{tcilatex}
\begin{document}

\title{Throwing bridges: where and how can classical and quantum views be
connected? }
\author{Claudio Garola \\
Department of Mathematics and Physics, University of Salento\\
Via Arnesano, 73100 Lecce, Italy\\
e-mail: garola@le.infn.it}
\maketitle

\begin{abstract}
We have proposed in several recent papers a critical view of some parts of
quantum mechanics (QM) that is methodologically unusual because it rests on
analysing the language of QM by using some elementary but fundamental tools
of mathematical logic. Our approach proves that some widespread beliefs
about QM can be questioned and establishes new links with a classical view,
which is significant in the debate on the interpretations of QM. We propose
here a brief survey of our results, highlighting their common background. We
firstly show how quantum logic (QL) can be embedded into classical logic
(CL) if the embedding is required to preserve the logical order and not the
algebraic structure, and also how QL can be interpreted as a pragmatic
sublanguage within a pragmatic extension of CL. Both these results challenge
the thesis that CL and QL formalize the properties of different and
incompatible notions of truth. We then show that quantum probability admits
an epistemic interpretation if contextuality is taken into account as a
basic constituent of the language of QM, which overcomes the interpretation
of quantum probability as ontic. Finally, we show that the proofs that QM is
a contextual theory stand on a supplementary epistemological assumption that
is usually unnoticed and left implicit. Dropping such assumption opens the
way, at least in principle, to non-contextual interpretations of QM.\bigskip 

\textbf{Keywords} Quantum logic\textperiodcentered Quantum
pragmatics\textperiodcentered Quantum probability\textperiodcentered
Noncontextual\textperiodcentered interpretations of quantum mechanics
\end{abstract}

\section{Introduction}

After almost one century after the birth of quantum mechanics (QM), the
debate on the interpretation of this theory is still alive (see, e.g.,
Schlosshauer et al., 2013). At variance with classical mechanics (CM), QM
exhibits indeed some features that make it hard to construct a consistent
picture of the physical world. In particular, a highly counterintuitive
feature of QM is \textit{nonlocality}, following (according to the standard
interpretation of QM) from the existence of entangled states, which
nevertheless are considered as important resources in quantum information
theory. Based on a critical analysis of the theorems aiming at proving
nonlocality, we proposed in some previous papers (see, e.g., Garola, 2015;
Garola et al., 2016) a falsifiable generalization of QM that embodies the
mathematical formalism of QM but introduces a reinterpretation of quantum
probabilities able to avoid nonlocality. However, we will not insist on this
proposal here, nor consider the numerous interpretations that have been
advanced and their variants. We would rather focus on the results obtained
in some further papers, in which we made a critical analysis of some typical
issues in QM (Garola, 2017; Garola, 2018; Garola and Persano, 2014; Garola
and Sozzo, 2010, 2013). Indeed, those results are significant, in our
opinion, if one wants to choose or to construct an interpretation of QM, for
they show that some widespread beliefs about QM can be questioned.

Let us list the issues that are considered in the papers quoted above.

(i) The \textit{contextuality} of QM, which implies that the values of the
observables of a physical system do not pre-exist to their measurements.
This is a basic notion in the standard interpretation of QM, but a formal
proof of it has been given for the first time by the famous Bell's (1966)
and Kochen-Specker's (1967) theorems. It can be rephrased by saying that the
properties of a quantum physical system are non-objective, in the sense that
assuming that they are either possessed or not possessed by the system
independently of any measurement leads to contradictions.

(ii) The controversial role of \textit{(standard, sharp) quantum logic} (QL)
in QM. Some authors maintain indeed that QL formalizes a new way of
reasoning introduced by QM, as this theory would determine a notion of
quantum truth, different and incompatible with the classical notion of truth
(see, e.g., R\'{e}dei, 1958; Dalla Chiara et al., 2004). Other authors
uphold instead that QL is just a part of the mathematical apparatus of QM
and has nothing to do with logic (see, e.g., Aerts, 1999).

(iii) The interpretation of \textit{quantum probability}, whose mathematical
structure is different from that of classical probability. Indeed, quantum
probability is usually maintained to be non-epistemic, in the sense that it
cannot be interpreted as expressing a (indirect) measure of our ignorance of
values of observables pre-existing to measurements.

Issue (ii) is discussed in (Garola, 2008; Garola, 2017; Garola and Sozzo,
2014) and issue (iii) in (Garola, 2018). Both treatments accept the standard
view about non-objectivity of properties. Nevertheless, we obtain some
results that question standard beliefs, as anticipated above. Indeed, when
considering QL in (Garola, 2008; Garola and Sozzo, 2014), we show that it
can be embedded into classical logic (CL) if the embedding is required to
preserve the logical order and not the algebraic structure; moreover, we
prove in (Garola, 2017) that QL can be interpreted as a pragmatic
sublanguage within a pragmatic extension of CL. Both these results challenge
the thesis that CL and QL formalize the properties of different and
incompatible notions of truth. Furthermore, when considering quantum
probability in (Garola, 2018), we show that it admits an epistemic
interpretation if contextuality is taken into account as a basic constituent
of the language of QM, which overcomes the aforesaid non-epistemic
interpretation of quantum probability.

Finally, issue (i) is discussed in (Garola and Sozzo, 2010; Garola and
Persano, 2014). Here we show that the proofs that QM is a contextual theory
stand on a supplementary epistemological assumption that is usually
unnoticed and left implicit. Dropping such assumption opens the way, at
least in principle, to non-contextual interpretations of QM, which are
usually excluded on the basis of the conviction that contextuality is
\textquotedblleft mathematically proven\textquotedblright\ by Bell's and
Kochen-Specker's theorems.

In this paper we present the above arguments in a unified perspective,
avoiding technical details but trying to make clear the physical, logical
and epistemological reasons underlying them. In particular, we aim to show
that our procedures exemplify a somewhat unusual methodology in physics,
because they are based on analysing the language of QM by means of some
elementary but fundamental tools of mathematical logic, which also leads to
establish some new links with classical views. To this end, we anticipate
some epistemological and linguistic preliminaries in Section 2 and then
summarize the main lines of our analysis of the issues listed above in
Section 3 (issue (ii)), Section 4 (issue (iii)) and Section 5 (issue (i); in
this case our treatment is also refined with respect to its original
presentations), insisting on their common background. We hope that this
presentation may constitute a useful contribution to the debate on the
possible interpretations of QM.

\section{Epistemological and linguistic preliminaries}

As we have anticipated in Section 1, we recollect and summarize in this
section some notions that are well known to epistemologists (Section 2.1) or
to physicists (Sections 2.2 and 2.3). Our presentation, however, is somewhat
original, for it mainly focuses on the languages of the theories that are
considered (CM and QM), thus constituting a background for our arguments in
Sections 3, 4 and 5.

\subsection{The received view}

According to the \textit{standard epistemological conception}, or \textit{%
received view} (see, e.g., Braithwaite, 1953; Nagel, 1961; Hempel, 1965;
Carnap, 1966), a fully-developed physical theory $\mathcal{T}$ is in
principle expressible by means of a metalanguage in which a \textit{%
theoretical language} $L_{T}$ and an \textit{observational language} $L_{O}$
can be distinguished. The theoretical apparatus of $\mathcal{T}$ includes a 
\textit{mathematical structure} expressed by means of $L_{T}$ and, usually,
an \textit{intended interpretation}, namely a \textit{direct} and \textit{%
complete} physical model of the mathematical structure. Such an
interpretation is often anticipated by the choice of the nouns of the
theoretical terms and it is not indispensable in principle, but plays a
fundamental role in the intuitive comprehension, justification and
development of the theory\ (think, e.g., to the trajectories of point-like
particles in CM or to the interpretation of electromagnetic fields as
waves). The observational language $L_{O},$\ instead, is interpreted via 
\textit{assignment rules} on an empirical domain, hence it has a semantic
interpretation. The two languages are connected by \textit{correspondence}
(or \textit{epistemic})\textit{\ rules} $R_{C}$ that establish complex and
sometimes problematic relations between $L_{O}$ and $L_{T}$ (e.g., a
self-adjoint operator may represent many different but physically equivalent
measuring devices in QM), so that the correspondence together with the
assignment rules provide an \textit{empirical interpretation} of the
mathematical structure. This interpretation generally is \textit{indirect},
in the sense that there are theoretical entities that are connected with the
empirical domain only via \textit{derived} theoretical entities, and \textit{%
incomplete}, in the sense that only limited ranges of values of the
theoretical entities are interpreted (e.g., self-adjoint operators
correspond in QM to measuring apparatuses whose outcomes match the
eigenvalues of the operators only in finite intervals of the real axis).

The received view was criticized by some authors (see, e.g., Kuhn, 1962;
Feyerabend, 1975) and is nowadays maintained to be outdated by several
scholars. Nevertheless, we deem that its basic ideas are still
epistemologically relevant and may greatly help to single out the
fundamental differences between CM and QM. In particular, this view led us
to focus our attention on the languages of physical theories, suggesting to
explore their similarities and differences by analysing their syntax and
semantics. We review in the following sections several results that we have
obtained following that suggestion, some of which are quite unexpected and
challenge well established beliefs.

We add that in the standard language of physical theories the distinctions
introduced by the received view are usually overlooked, and the various
linguistic components are mixed together (e.g., the term \textquotedblleft
observable\textquotedblright\ may denote in QM a self-adjoint operator on a
Hilbert space $\mathcal{H}$, and in this sense it belongs to $L_{T}$, but
also a physical entity associated with a set of measurement procedures, and
in this sense it belongs to $L_{O}$; the term \textquotedblleft
state\textquotedblright\ may denote a vector of $\mathcal{H}$, but also a
physical entity associated with a set of preparing procedures; etc.). Only a
rational reconstruction of the language of a theory can lead to distinguish
clearly the various elements that occur in it according to the received
view. We, however, will not deal with this issue in this paper. It will be
indeed sufficient for our aims to formalize some simple sublanguages of the
languages of CM and QM which exhibit strong similarities and differences.

\subsection{The language of classical mechanics}

In the standard language of CM the notions of \textit{material body} and 
\textit{physical quantity} play a basic role, both at a theoretical and at
an observational level. According to the received view (Section 2.1), the
terms \textquotedblleft material body\textquotedblright\ and
\textquotedblleft physical quantity\textquotedblright\ belong to the
observational language $L_{O}$ of CM. The former refers to a set of elements
(\textit{material bodies}) of the empirical domain that are represented in
the theoretical language $L_{T}$ of CM by the elements of an abstract set,
still called material bodies by abuse of language. The latter refers to a
set of entities (\textit{physical quantities}) of the empirical domain, each
of which consists of a set of (exact) \textit{measurement procedures} and is
represented in $L_{T}$ by a function that takes values on material bodies,
still called physical quantity by abuse of language; then, every measurement
procedure belonging to a physical quantity, when activated on a material
body, performs a \textit{measurement} whose outcome yields the value of the
physical quantity on the body.

Based on the notions mentioned above, the derived notions of \textit{property%
} and \textit{state} are introduced. Also the terms \textquotedblleft
property\textquotedblright\ and \textquotedblleft state\textquotedblright
belong to $L_{O}$. The former refers to a set of entities (\textit{properties%
}) of the empirical domain, each of which consists of a set of (exact) 
\textit{dichotomic} measurement procedures and is represented in $L_{T}$ by
a pair of the form $(A,\Delta )$, still called property by abuse of
language, where $A$ is a physical quantity and $\Delta \in \mathcal{B}(R)$
is a Borel set of the real line (hence the physical quantity $A$ is
bijectively associated with the family $\{(A,\Delta )\}_{\Delta \in \mathcal{%
B}(R)}$ of properties); then, one says that a material body $a$ \textit{%
possesses} the property $E$ represented by $(A,\Delta )$ if and only if (%
\textit{iff} in the following) the value that $A$ takes on $a$ belongs to $%
\Delta $, so that every dichotomic measurement procedure associated, via $%
L_{O}$, with $E$, when activated on a material body $a$, performs a
measurement whose outcome tells us whether $a$ possesses $E$. The latter
term refers to a set of entities (\textit{states}) of the empirical domain,
each of which consists of a set of \textit{preparation procedures} and is
represented in $L_{T}$ by a set of properties, still called state by abuse
of language; then, every preparation procedure belonging to a state $S$,
when activated, prepares material bodies sharing the set of properties
representing $S$.

States and properties play a fundamental role in our analysis. Indeed, the
statement that a material body $a$ has been prepared by a preparation
procedure associated with the state $S$ (briefly, \textquotedblleft $a$ is
in the state $S$\textquotedblright ) can be formalized by an elementary
sentence $S(a)$ of predicate logic. Analogously, a statement of the form
\textquotedblleft $a$ possesses the property $E$\textquotedblright\ can be
formalized by $E(a)$.

The semantic rules assigning truth values to the sentences of $L_{O}$ are
now crucial. Indeed, truth assignments are made in the language of CM
according to the classical theory of truth as correspondence, as
reconstructed by Tarski (1944, 1956). When considering elementary sentences,
truth values (\textit{true}/\textit{false}) are assigned by fulfilling
Tarski's truth condition (exemplified by the famous statement
\textquotedblleft `the snow is white' is true iff the snow is
white\textquotedblright ). This is apparent if one refers to the standard
geometrical model of the mathematical apparatus, in which a physical system
is represented by a phase space and properties and states by subsets and
points, respectively, of the phase space. Indeed, according to this model, a
property $E$ is possessed by the individual object $a$ in the state $S$,
that is, the sentence $E(a)$ is true, iff the point representing $S$ belongs
to the subset representing $E$. This rule, however, is not semantically
neutral as such an assignment implies an assumption of \textit{objectivity
of properties}, which can be stated as follows.

O. Any property of a material body $a$ is possessed or not possessed by $a$
independently of any measurement.

Usually, assumption O is not stated explicitly but is implicit in the
language of CM. It implies in particular that the measurement procedures
associated with $E$, when performed on the empirical object $a$, check (or
reveal, if unknown) the truth value of $E(a)$, which does not depend on the
procedures themselves and pre-exists to them: hence, all procedures must
yield the same result.

When considering complex sentences that can be formalized by introducing
classical logical connectives as $\lnot $ (\textit{not}),\textit{\ }$\wedge $%
\ (\textit{and}),\textit{\ }$\vee $\textit{\ }(\textit{or}), etc., and then
connecting elementary sentences of the form $E(a)$ or $S(a)$ by means of
these connectives, Tarski's theory requires that truth values be assigned by
recursive rules such that the truth value of any complex formula depends
only on the truth values of its elementary subformulas: i.e., every truth
assignment is \textit{T-functional} (which implies that the meaning of the
logical connectives is independent of the empirical interpretation of $L_{O}$%
).

It is now important to observe that assumption O is coupled in CM with an
assumption of \textit{compatibility of properties}, which can be stated as
follows.

C. The measurement procedures associated with different properties can be
performed conjointly.

Assumption C implies indeed that the truth values of the complex sentences
of the kind considered above can be checked by checking the truth values of
all elementary sentences that occur in them and by using standard truth
rules in classical predicate logic. In particular, a sentence of the form $%
S(a)$ can be seen as logically equivalent to a conjunction of elementary
sentences, each of which states that a possesses a given property. Hence,
also the truth value of $S(a)$ can be checked (or revealed) by means of
measurements

\subsection{The language of quantum mechanics}

It is well known that QM has been considered a problematic theory since its
birth, and that many \textquotedblleft interpretations\textquotedblright\ of
it have been proposed. We avoid dealing with this issue here, and refer only
to the \textquotedblleft standard interpretation\textquotedblright\ (also
\textquotedblleft Copenhagen interpretation\textquotedblright ) of QM,
maintaining that QM deals with individual examples of quantum physical
systems (briefly, \textit{individual}, or \textit{physical}, \textit{objects}%
) and their properties (we remind that this option is classified as \textit{%
realistic} by some scholars; see, e.g., Busch et al., 1996).

When considering the language of the standard interpretation of QM and
comparing it with the language of CM, it is apparent that the basic notions
are similar, while their relations and interpretations are deeply different.
To be precise, the notions of state and property can be defined by replacing
the notions of material body and physical quantity with the notions of
individual object and observable, respectively, in the definitions
introduced in the language of CM. The mathematical apparatus of QM is
obviously very different from the mathematical apparatus of CM, but
sentences as \textquotedblleft the individual object $a$ is in the state $S$%
\textquotedblright\ or \textquotedblleft the individual object $a$ possesses
the property $E$\textquotedblright\ still occur in the language of QM and
can be formalized by $S(a)$ and $E(a)$, respectively. Hence classical
connectives as $\ \lnot $,\textit{\ }$\wedge $,\textit{\ }$\vee $, etc., can
be formally introduced to construct complex sentences also in QM.
Nevertheless, the semantic rules of classical logic are not adequate to
match the empirical domain that the language of QM aims to describe. It is
well known indeed that in Bohr's holistic view (see, e.g., Bohr, 1958) or in
Heisenberg's distinction between \textquotedblleft
potential\textquotedblright\ and \textquotedblleft actual\textquotedblright\
properties (see, e.g., Heisenberg, 1958), assumption O in Section 2.2
(objectivity of properties) does not hold, as the outcome of any measurement
of a given property on a given individual object depends on the choice of
the (macroscopic) measurement procedure. This epistemological view, that has
of course an enormous impact on our conception of the physical world and has
given rise to a huge literature, is maintained to be \textquotedblleft
mathematically proven\textquotedblright\ in QM because of several
\textquotedblleft no-go\textquotedblright\ theorems, the most important of
which are Bell's (1964, 1966) and Kochen-Specker's (1967). While we think
that the proofs of these theorems depend on some implicit epistemological
assumptions that may be questioned, thus opening the way to different
interpretations of QM (see Section 5), for the moment we maintain the
standard view that QM is a \textit{contextual} theory (Bell, 1966; Kochen
and Specker, 1967) and that contextuality occurs also at a distance (\textit{%
nonlocality}: Bell, 1964).

Contextuality and nonlocality have numerous puzzling consequences. We aim to
deal with the following in the present paper.

(i) We have seen in Section 2.1 that the intended interpretation of the
theoretical language $L_{T}$ of a physical theory is not logically necessary
but plays a fundamental role in the intuitive comprehension of the theory.
But contextuality implies, in the case of QM, that it is impossible to
supply an intended interpretation (that is, a complete and direct physical
model) of $L_{T}$ in which individual objects are represented together with
their properties. Thus, no intuitive picture of the physical world can be
given, which may explain why a prominent physicist as Feynmann said
\textquotedblleft \ldots I think I can safely say that nobody understands
quantum mechanics\textquotedblright\ (Feynmann, 1964). The famous duality
between particle and wave models for QM finds its roots in that
impossibility.

(ii) When considering a composite quantum system, nonlocality implies that
measuring a property of a part of the system may instantaneously actualize a
property of another part, even if the latter is far away from the former.
This \textquotedblleft spooky action at a distance\textquotedblright , as
Einstein classified it, is commonly accepted nowadays as a consequence of
entanglement, but strongly clashes with our intuitive conception of space
and time (EPR paradox) even if it does not imply any transmission of
information.

(iii) The necessity of giving up assumption O in QM implies that truth
values can be assigned to sentences of the form $E(a)$ only by referring to
specific measurement procedures. But, then, it turns out that different
measurement procedures associated with the same property may yield different
outcomes when activated on a given individual object, and that the same
measurement procedure, when activated on different individual objects in the
same state, may also yield different outcomes. Moreover, it may occur that
procedures associated with different properties are incompatible, that is,
they cannot be performed conjointly, so that also assumption C in Section
2.2 does not hold in QM. It follows that, even if a formal language with all
elementary sentences of the form $S(a)$ and $E(a)$ and connectives $\lnot $,%
\textit{\ }$\wedge $,\textit{\ }$\vee $, etc., is constructed, no classical
semantics can be defined on it in such a way that it formalizes a proper
sublanguage of the language of QM. This remark has suggested the idea that a
new, non-classical, notion of truth (\textit{quantum truth}) is determined
by QM, hence a new logic, i.e. QL. The research on QL started indeed with a
famous paper by Birkhoff and von Neumann (1936), which gave rise to an
enormous literature, and many scholars maintain that QL formalizes the basic
language of QM (see, e.g., R\'{e}dei, 1958; Dalla Chiara et al., 2004).

(iv) Classical probability, whose mathematical structure is formalized by
Kolmogorov's probability theory, be its \textquotedblleft
interpretation\textquotedblright\ logical, or frequentist, or subjectivist,
is usually maintained to be \textit{epistemic}, i.e., to express our
incomplete knowledge of the empirical world (hence of the truth values of
the sentences of the language that describes it). But an elementary sentence
of the observational language of QM generally has no truth value before a
measurement, hence quantum probability cannot be considered as an (indirect)
measure of our ignorance. Therefore it is often classified as \textit{ontic}%
, and many scholars maintain that it constitutes an intrinsic feature of the
physical world. Moreover, quantum probability is defined on the set of 
\textit{propositions} of QL, which is an orthomodular non-Boolean lattice,
hence it does not satisfy Kolmogorov's axioms.

The above consequences of contextuality and nonlocality seem to establish
the incompatibility of the classical view of the physical world with the
quantum view. As anticipated in Section 1, however, a deeper analysis shows
that the classical view can be extended in different directions, which
allows to establish some unexpected connections with QL and quantum
probability (Sections 3 and 4, respectively). Moreover, we intend to show
that, at least in principle, a new interpretation of QM recovering O cannot
be excluded (Section 5).

\section{Bridging classical and quantum logic}

We have seen in Section 2.3 that some authors maintain that QM implicitly
determines a new notion of truth, hence a new logic (QL), which is
incompatible with CL. Other authors (see, e.g., Aerts, 1999) uphold instead
that QL simply formalizes empirical relations among properties in QM, not a
new logic.

We have shown in several previous papers that an order structure isomorphic
to QL can be singled out inside CL (Garola, 2008; Garola and Sozzo, 2013),
and also that CL can be pragmatically extended in such a way that QL can be
seen as a part of that pragmatic extension (Garola, 2017). Both procedures
establish bridges between the two logics. The former allows a formal
embedding of QL into CL preserving a new order (physical preorder) that is
implied by the standard logical order but is generally weaker than it (it is
well known instead that no embedding of QL into CL preserving the algebraic
structure is possible because the algebraic structures of the two logics are
different). The latter is innovative and requires a generalization of the
pragmatic language introduced by ourselves together with another author
several years ago to supply a pragmatic interpretation of intuitionistic
logic (Dalla Pozza and Garola, 1995). The two procedures have different
merits but can be interconnected by showing that they are based on the same
perspective. We resume both of them here focusing mainly on their general
features.

\subsection{Embedding quantum logic into classical logic}

We will present our embedding by referring to our more recent paper on this
issue (Garola and Sozzo, 2013). The starting point is the construction of a
\textquotedblleft concrete logic\textquotedblright , proceeding as follows.

(i) Let $\mathcal{T}$ be a physical theory in which the notions of physical
object, property and state are introduced. A classical formal language $L(x)$
is constructed that is intended to express basic relations in $\mathcal{T}$
(hence $L(x)$ is a sublanguage of the theoretical language of $\mathcal{T}$%
). The syntax of $L(x)$ consists of two parts: firstly, a logical
vocabulary, or \textit{alphabet}, which contains two disjoint sets of
monadic predicates (called the set $\mathcal{E}$ of \textit{properties} and
the set $\mathcal{S}$ of \textit{states}, see Section 2.2), standard
connectives $\lnot $,\textit{\ }$\wedge $,\textit{\ }$\vee $, $\rightarrow ,$
an individual variable $x$ and parentheses; secondly, \textit{standard
formation rules}, which define a set $\psi (x)$ of elementary and complex 
\textit{well-formed formulas} (\textit{wffs}). The (formal) semantics of $%
L(x)$ consists of a \textit{universe} $\mathcal{U}$ of \textit{physical
objects}, a set $\Sigma $ of \textit{interpretations of the variable} $x$,
and, for every $\sigma \in \Sigma $, a \textit{truth assignment} $\nu
_{\sigma }$ that associates a truth value (\textit{t}/\textit{f}, where 
\textit{t} stands for \textit{true} and \textit{f} for \textit{false}) with
every wff of $\psi (x)$, following classical truth rules. The logical
preorder $<$ and the logical equivalence $\equiv $ are then defined on $\psi
(x)$ in a standard way, i.e., by setting, for every $\alpha (x)$, $\beta
(x)\in \psi (x)$, $\alpha (x)<\beta (x)$ iff, for every $\sigma \in \Sigma $%
, $\nu _{\sigma }(\beta (x))=t$ whenever $\nu _{\sigma }(\alpha (x))=t$, and 
$\alpha (x)\equiv \beta (x)$ iff $\alpha (x)<\beta (x)$ and $\beta
(x)<\alpha (x)$.

(ii) A subset $\phi (x)\subset \psi (x)$ is introduced whose elements are
all wffs of $\psi (x)$ in which no symbol of state occurs. Based on the
classical notion of truth, a derived notion of C-truth is defined on $\phi
(x)$ by stating that a wff $\alpha (x)\in \phi (x)$ is \textit{certainly true%
} (\textit{certainly false}) in a state $S$ iff, for every $\sigma \in
\Sigma $, $\nu _{\sigma }(S(x))=t$ implies $\nu _{\sigma }(\alpha (x))=t$\ ($%
\nu _{\sigma }(\alpha (x))=f$).

(iii) A \textit{physical preorder} $\prec $ is defined on $\phi (x)$ by
setting, for every $\alpha (x)$, $\beta (x)\in \phi (x)$, $\alpha (x)\prec
\beta (x)$ iff, for every $S\in \mathcal{S}$, $\alpha (x)$ certainly true in 
$S$ implies $\beta (x)$ certainly true in $S$. Moreover, a \textit{physical
equivalence} $\approx $\ is defined on $\phi (x)$ by setting, for every $%
\alpha (x)$, $\beta (x)\in \phi (x)$, $\alpha (x)\approx \beta (x)$ iff $%
\alpha (x)\prec \beta (x)$ and $\beta (x)\prec \alpha (x)$. Then, it can be
proved that $<$ implies $\prec $ and $\equiv $ implies $\approx $.

(iv) A notion of \textit{verification} is defined in $L(x)$ by considering 
\textit{verifiable} (according to $\mathcal{T}$) all wffs of $\phi (x)$ that
are logically equivalent to elementary wffs of $\phi (x)$.

(v) Let $\phi _{V}(x)$ be the subset of all verifiable wffs of $\phi (x)$,
and let $\mathcal{T}$ induce a \textit{weak orthocomplementation }$^{\bot }$
on $(\phi _{V}(x),\prec )$ (i.e., a mapping of $\phi _{V}(x)$ into itself
such that, for every $\alpha (x)\in \phi _{V}(x)$, $(\alpha (x))^{\bot \bot
}\approx \alpha (x)$, and for every $\alpha (x)$, $\beta (x)\in \phi _{V}(x)$%
, $\alpha (x)\prec \beta (x)$ implies $(\beta (x))^{\bot }\prec (\alpha
(x))^{\bot }$). Then, the structure $(\phi _{V}(x),\prec ,^{\bot })$ is the 
\textit{concrete logic} associated with $\mathcal{T}$.

\bigskip 

When considering CM, the phase space representation of physical systems
shows that C-truth coincides with classical truth on $\phi (x)$. Moreover,
all wffs of $\phi (x)$ are verifiable, at least in principle, according to
CM, so that one can set $\phi _{V}(x)=\phi (x)$. Hence, the concrete logic
of CM has the structure of a classical logic. It is important to note,
however, that there may be physical theories of macroscopic systems in which
the testability criteria are restricted, so that $\phi _{V}(x)$ is a proper
subset of $\phi (x)$. In these cases, one obtains concrete logics whose
algebraic structure may be very different from the structure of Boolean
lattice that characterizes CL. Examples of these logics are provided
(without referring to $L(x)$) by the macroscopic systems considered by Aerts
to show that quantum logical structures can be obtained in suitably chosen
macroscopic domains (see, e.g., Aerts, 1999; Garola and Sozzo, 2013).

Aerts' results are relevant because they falsify the belief that QL
characterizes QM, as they show that quantum structures can be obtained also
in a classical framework. We attain the same conclusion in a generalized
form by considering $L(x)$. Indeed, the classical semantics introduced in $%
L(x)$ implies that $L(x)$ cannot be seen as a sublanguage of the language of
QM because of the reasons expounded in Section 2.3. Nevertheless, one can
show that, if $\phi _{V}(x)$ satisfies suitably chosen axioms, then $(\phi
_{V}(x),\prec ,^{\bot })$ is a structure isomorphic to QL (up to an
equivalence relation). This conclusion implies that QL can be embedded into
CL and interpreted as a structure formalizing the properties of a notion of 
\textit{true with certainty} within $\phi _{V}(x)$\ rather than the
properties of a notion of quantum truth,\ alternative to the classical
notion of truth. This interpretation of QL is also supported by some
traditional approaches to quantum physics, as Piron's (1976), in which the
orthomodular structure of the set of quantum \textquotedblleft
propositions\textquotedblright\ is recovered standing on the notion of true
with certainty.

\subsection{Recovering quantum logic within a pragmatic extension of
classical logic}

We have discussed with another author in a previous paper (Dalla Pozza and
Garola, 1995) how to pragmatically extend classical propositional logic in
such a way that intuitionistic propositional logic can be embedded into the
pragmatic part of the extension and reinterpreted as the logic formalizing
the properties of the metalinguistic notion of constructive logical proof.
Such a reinterpretation has a deep philosophical meaning. Indeed, it shows
that one can avoid introducing an intuitionistic notion of truth, different
and incompatible with classical truth, by adopting the perspective of 
\textit{global pluralism}, according to which many different logical systems
can coexist without being in competition, for they formalize the properties
of different metalinguistic notions (Haak, 1974, 1978; Garola, 1992).

Our pragmatic extension is based on a distinction that goes back to Frege
(1893) and that has given rise to a huge literature in linguistic and
philosophical studies: that is, the distinction between the sentences of a
language, which can be true or false, and the \textit{assertions} of a
speaker who commits himself to the truth of the sentences he is uttering.
Indeed, an assertion has not a truth value, but it can only be \textit{%
justified} (if a proof exists that the asserted sentence is true) or \textit{%
unjustified} (if no proof exists that the asserted sentence is true).
Bearing in mind this distinction, we firstly introduce a standard classical
propositional logic $\mathcal{L}$, whose formulas we call \textit{radical
formulas}. Then we construct an extension $\mathcal{L}^{P}$ of $\mathcal{L}$
by adjoining a new category of \textit{logical-pragmatic signs}, which
contains an \textit{assertion sign} $\vdash $ and \textit{pragmatic
connectives} $N$, $K$, $A$, $C$ and $E$, to the alphabet of $\mathcal{L}$.
By using this extended vocabulary, new formation rules are introduced which
recursively define a set of elementary and complex assertive formulas of $%
\mathcal{L}^{P}$. Every elementary assertive formula consists of a radical
formula preceded by the assertion sign, and every complex assertive formula
consists of elementary assertive formulas connected by pragmatic
connectives. Then, \textit{pragmatic rules} are introduced which specify the
conditions that must be fulfilled whenever a \textit{pragmatic evaluation
function} on the set of all assertive formulas of $\mathcal{L}^{P}$ is given
which assigns a \textit{justification value} (\textit{justified}/\textit{%
unjustified}) to every assertive formula. Such a value is defined in terms
of an informal notion of proof, and the assignment of a justification value
to an assertive formula of $\mathcal{L}^{P}$ depends on the semantic
assignment of truth values to the radical subformulas that occur in it.
Moreover, pragmatic evaluation functions are not \textit{J-functional},
i.e., the justification value of a complex assertive formula generally does
not depend only on the justification values of its elementary assertive
subformulas because of the features of the informal notion of proof that is
adopted (which implies an intuitionistic-like behaviour of the pragmatic
connectives).

The language $\mathcal{L}^{P}$ is original from several points of view: in
particular, because of the introduction of pragmatic connectives and
pragmatic evaluation functions. It provides a pragmatic extension of
classical propositional logic that allows us to embed intuitionistic
propositional logic in it, as we have just seen. But it may have also a more
general role: whenever the informal notion of proof is suitably specified,
it provides a framework in which non-classical logics (and not only
intuitionistic logic) can be embedded as fragments of its pragmatic part.
Hence it was natural for us to wonder, in particular, whether this procedure
could apply to QL.

The idea of recovering QL within $\mathcal{L}^{P}$, however, meets a serious
difficulty from the very beginning. Indeed, we have already seen in the
previous sections that the language of QM cannot bear a classical semantics.
Hence a generalization of the original language $\mathcal{L}^{P}$ is needed
to implement that idea. We introduce such a generalization in the recent
paper mentioned at the beginning of this section (Garola, 2017), retaining
the syntactic apparatus of $\mathcal{L}^{P}$ and its pragmatic rules, but
assuming that the truth assignments on radical formulas can be \textit{%
partial}, i.e., such that not all elementary and complex radical formulas
have a truth value (but classical truth rules for assigning truth values to
complex radical formulas are preserved, so that each truth assignment is
T-functional in the sense specified in Section 2.2). This generalization
makes our pragmatic language ($\mathcal{L}_{G}^{P}$ in the following)
suitable for dealing not only with the standard interpretation of QM, but
also with other interpretations, as the \textit{modal interpretations} or
the \textit{objective interpretation} proposed by ourselves on the basis of
our criticism of Bell's and Kochen-Specker's theorems (see Section 5). Hence
we can single out a fragment $\mathcal{L}_{GQ}^{P}$ of $\mathcal{L}_{G}^{P}$
that we call \textit{quantum pragmatic language} and introduce a physical
(intended) interpretation of it that specifies the notion of proof as the
notion of \textit{empirical proof} in QM and supplies $\mathcal{L}_{GQ}^{P}$
with a semantics (which depends on the interpretation of QM that has been
chosen) and a pragmatics (which does not depend on the choice of the
interpretation of QM), in agreement with the rules mentioned above. It is
then easy to show that $(\psi _{AQ},\prec )$, where $\psi _{AQ}$ denotes the
set of all assertive formulas of $\mathcal{L}_{GQ}^{P}$ and $\prec $ denotes
the preorder induced on $\psi _{AQ}$ by the set of all pragmatic evaluation
functions, is isomorphic to QL, considered as an order structure, up to an
equivalence relation. This result implies that the connectives of QL can be
identified, up to an equivalence relation, with (primitive or derived)
pragmatic connectives of $\mathcal{L}_{GQ}^{P}$. Hence QL can be interpreted
as a structure formalizing the properties of the notion of \textit{empirical
justification} in QM rather than of a notion of quantum truth.

The conclusion above matches the conclusion obtained in Section 3.1 if the
notion of verification is replaced by the equivalent notion of empirical
justification. Moreover, whenever the standard interpretation of QM is
adopted, the truth values (\textit{true}/\textit{false}) assigned to the
radical formulas match the values of C-truth assigned in Section 3.1 to the
formulas of $\phi (x)$. But in Section 3.1 we have seen that QL is proven to
be isomorphic (up to an equivalence relation) to an order structure embedded
into a classical predicate logic. In the framework presented in this
section, instead, QL is embedded into a pragmatic extension of a classical
propositional logic, and the embedding is independent of the interpretation
of QM that is adopted. This makes the relation between QM and CL more
intuitive. The conclusion that QL does not characterize QM is instead made
more evident by the approach in Section 3.1. In both cases, however, a
bridge between the classical and the quantum views is thrown.

\section{An epistemic interpretation of quantum probability}

We have recalled in Section 2.3 that quantum probability is maintained to be
radically different from classical probability by many scholars. Indeed,
besides having a non-classical mathematical structure, it would not admit an
epistemic interpretation. This view, however, can be questioned. We have
shown indeed in a recent paper (Garola, 2018) that quantum probability can
be recovered, under reasonable assumptions that take into account
contextuality, as a derived notion in a classical probabilistic framework.
This result throws another bridge between the classical and the quantum
views, and implies that quantum probability can be considered epistemic, at
variance with the standard view.

Our treatment in the paper mentioned above starts from a simple remark. The
basic language of QM considered in Section 2.3, which is strongly influenced
by the language of CM, makes no reference to contextuality, which is
introduced as a complex theoretical notion that can be expressed only by a
higher order language. But contextuality is a fundamental feature of QM,
which suggests that it should enter its language from the very beginning as
an essential notion associated with the notion of property. Moreover, one
can suspect that such a change in the basic language of QM could make it
possible to introduce a classical semantics on it, avoiding the problems
mentioned in Section 2.3.

A hint on how to implement the above suggestions is given by our comments on
the language of QM in Section 2.3. We can indeed retain all elementary
sentences of the form \textquotedblleft the individual object $a$ is in the
state $S$\textquotedblright , formalized by $S(a)$, and replace every
elementary sentence of the form \textquotedblleft the individual object $a$
possesses the property $E$\textquotedblright\ with the sentence
\textquotedblleft the individual object $a$ possesses the property $E$ in
the context $C$\textquotedblright\ or, briefly, \textquotedblleft the
individual object $a$ possesses the contextual property $E_{C}$%
\textquotedblright . A sentence of this kind could then be formalized by $%
E_{C}(a)$ in the new basic language of QM.

Of course, the proposal above requires specifying an empirical
interpretation of the context $C$. At first sight, one could think of $C$ as
the macroscopic measurement context determined by a measurement procedure
associated with a property $E$. But it is well known that QM predicts that
performing a given measurement on different individual objects in the same
state may yield different outcomes. This can be intuitively explained by
adopting a picture of the world according to which a microscopic world
underlies the macroscopic world of our everyday experience and by noticing
that there are two possible sources of randomness for the outcomes of a
measurement, as follows.

(i) When an individual object is prepared by activating a preparation
procedure associated with a state $S$ (see Sections 2.2 and 2.3), we control
only macroscopic variables, not the physical situation at a microscopic
level. Thus different individual objects produced by the preparation
procedure are not bound to possess the same contextual properties.

(ii) When performing a measurement, many \textit{microscopic contexts}
(which can be described, in principle, by QM itself) can be associated with
the (macroscopic) measurement procedure that is activated, and different
microscopic contexts that we cannot control may affect in different ways the
outcome of the measurement.

Remark (ii) suggests that $C$ should denote a microscopic context if we want
to assign a truth value to a sentence of the form $E_{C}(a)$. Bearing in
mind this suggestion, our first step in the paper mentioned above is
constructing a new language intended to serve as a basic elementary language
for theories belonging to a class $\mathbb{T}$ of theories in which the
notions of physical (or individual) object, state, property and context play
a fundamental role. This language is conceived as an extension of the
language $L(x)$ considered in Section 3.1, hence it will still be called $%
L(x)$ by abuse of language.

The syntax of $L(x)$ consists of two parts, i.e. an \textit{alphabet} and 
\textit{standard formation rules}. The former contains two disjoint sets $%
\mathcal{S}$ and $\mathcal{E}_{C}$ of monadic predicates, with $\mathcal{S}$
a set of \textit{states} and $\mathcal{E}_{C}=\{E_{C}=(E,C)|E\in \mathcal{E}%
,C\in \mathcal{C}\}$ a set of \textit{contextual properties} that is the
Cartesian product of a set $\mathcal{E}$ of \textit{properties}, each of
which is associated with a set $\mathcal{M}_{E}$ of \textit{measurement
procedures}, and a set $\mathcal{C}$ of \textit{microscopic contexts}
(briefly, $\mu $-contexts). Moreover, the alphabet of $L(x)$ contains
standard connectives $\lnot $,\textit{\ }$\wedge $,\textit{\ }$\vee $, an
individual variable $x$ and parentheses. The formation rules then define in
a standard way the set $\Psi (x)$ of elementary and complex \textit{%
well-formed formulas} (\textit{wffs}) of $L(x)$.

The (formal) semantics of $L(x)$ consists of a \textit{universe} $\mathcal{U}
$ of \textit{individual objects}, a set $\Sigma $ of \textit{interpretations
of the variable} $x$, and for every $\sigma \in \Sigma $, a \textit{truth
assignment} $\nu _{\sigma }$ that associates a truth value ($t$/$f$, where $t
$ stands for \textit{true} and $f$ for \textit{false}) with every wff of $%
\Psi (x)$ following classical truth rules (to be precise, an \textit{%
extension} $ext(\alpha (x))\subset \mathcal{U}$ of individual objects is
associated with every $\alpha (x)\in \Psi (x)$ in such a way that the set of
all extensions, ordered by set inclusion, is a Boolean lattice, and, for
every $\sigma \in \Sigma $, $\nu _{\sigma }(\alpha (x))=t$ iff $\sigma
(x)\in ext(\alpha (x))$). Finally, the \textit{logical preorder} $<$ and the 
\textit{logical equivalence} $\equiv $ are defined on $\Psi (x)$ in a
standard way.

It is now important to observe that the truth assignments on $\Psi (x)$ are
theoretical functions whose values generally cannot be checked by means of
measurements. Indeed, we cannot control the $\mu $-context underlying a
measurement, hence we cannot select a measurement whose outcome yields the
truth value of an elementary wff $E_{C}(x)$ when an interpretation $\sigma $
of the variable $x$ is given. Moreover, several different contextual
properties may occur in a complex wff $\alpha (x)\in \Psi (x)$ which refer
to different $\mu $-contexts. Hence different measurement procedures may be
required to check the truth value of $\alpha (x)$, which raises the problem
of their compatibility.

Our second step in (Garola, 2018) is suggested by remark (i). We construct
indeed a $\mu $\textit{-contextual probability structure} on $L(x)$ by
firstly introducing a probability measure on the Boolean lattice of all
extensions and then defining, for every pair $(\alpha (x),\beta (x))\in \Psi
(x)\times \Psi ^{+}(x)$ (where $\Psi ^{+}(x)$ is the subset of all wffs of $%
\Psi (x)$ whose extension has a non-zero probability measure), a $\mu $-%
\textit{contextual conditional probability} $p(\alpha (x)|\beta (x))\in
\lbrack 0,1]$ of $\alpha (x)$ given $\beta (x)$. Such a structure is
basically classical, hence $\mu $-contextual conditional probabilities admit
an epistemic interpretation. In other words, they can be considered as
indexes of our lack of knowledge of the (classical) truth assignments on $%
\Psi (x)$. However, these probabilities, as the truth assignments considered
above, generally cannot be checked by means of measurements. Indeed,
checking a $\mu $-contextual conditional probability would require
performing the same measurement on many individual objects, maintaining
under control both the preparation procedures at a microscopic level and the 
$\mu $-contexts underlying the measurement, which is impossible, as noticed
in remarks (i) and (ii).

Our third step in (Garola, 2018) is then looking for theoretical entities,
empirically interpreted via correspondence rules (see Section 2.1), whose
values can be checked. To this end we consider, for every property $E$, the
set $\mathcal{M}_{E}$ of all measurement procedures associated with $E$,
and, for every $M\in \mathcal{M}_{E}$, the macroscopic measurement context $%
C_{M}$ determined by $M$ and the set $\mathcal{C}_{M}$ of all $\mu $%
--contexts underlying $C_{M}$. Then, we associate a probability to each $\mu 
$-context $C\in \mathcal{C}_{M}$. This framework allows us to introduce a
binary relation $k$ of \textit{compatibility} on the set $\mathcal{E}$ of
all properties (the properties $E$ and $F$ are \textit{compatible} or,
briefly, $EkF$, iff they share at least one measurement procedure; hence $k$
is reflexive, symmetric, but generally not transitive). Moreover, it allows
us to select a subset of testable wffs of $\Psi (x)$(a wff $\alpha (x)$ is
testable iff all properties that occur in it are compatible and associated
with the same $\mu $--context or, conventionally, iff no contextual property
occurs in it, as in the case of the elementary wff $S(x)$) and to introduce
a notion of \textit{joint testability} on $\Psi (x)$ (the wffs $\alpha (x)$
and $\beta (x)$ are \textit{jointly testable} iff their conjunction is
testable). Whenever $\alpha (x)$ and $\beta (x)$ are jointly testable, they
share some measurement procedures. Choosing one of them, say $M$, and
assuming that, for every $C\in \mathcal{C}_{M}$, $\beta (x)\in \Psi ^{+}(x)$
(which will be understood in the following), we introduce the average of the 
$\mu $-contextual conditional probability of $\alpha (x)$ given $\beta (x)$
over all the $\mu $-contexts in $\mathcal{C}_{M}$. Whenever this average
does not depend on the choice of $M$, we denote it by $<p(\alpha (x)|\beta
(x))>$ and say that it is the \textit{mean conditional probability} of $%
\alpha (x)$ given $\beta (x)$.

Mean conditional probability is crucial in our approach. Indeed, we can now
focus our attention on the subclass $\mathbb{T}^{\prime }\subset \mathbb{T}$
of theories in which, for every pair $(\alpha (x),\beta (x))\in \Psi
(x)\times \Psi ^{+}(x)$ of jointly testable wffs, the foregoing condition on
the average is fulfilled and a mean conditional probability of $\alpha (x)$
given $\beta (x))$ is defined. Then, in every $\mathcal{T}\in \mathbb{T}%
^{\prime }$, we can assume that performing a measurement by activating a
measurement procedure $M$ shared by $\alpha (x)$ and $\beta (x)$ on a large
number of individual objects tells us the truth values of $\alpha (x)$ and $%
\beta (x)$ for different interpretations of the variable $x$, which allows
us to evaluate frequencies that we assume to check the mean conditional
probability $<p(\alpha (x)|\beta (x))>$. Intuitively, this assumption can be
justified by observing that both sources of randomness pointed out in
remarks (i) and (ii) underlie $M$. We therefore call the foregoing repeated
activation of $M$ on different individual objects a \textit{mean probability
measurement}.

Our general framework is thus completed. Its relevance and usefulness become
apparent when considering special cases. Bearing in mind the role of
properties and states in QM, let us consider, in particular, the elementary
wffs $E_{C}(x)$ and $S(x)$. Both these wffs are testable (the latter by
convention, as stated above). Moreover, they are obviously jointly testable,
and we can assume that $S(x)\in \Psi ^{+}(x)$ (the state $S$ would be
irrelevant if $ext(S(x))$ had probability measure zero). Thus, for every $%
E\in \mathcal{E}$ and $S\in \mathcal{S}$, we can consider the mean
conditional probability $<p(E_{C}(x)|S(x))>$ of $E_{C}(x)$ given $S(x)$,
which we briefly denote by $P_{S}(E)$ in the following. Hence, we can define
a preorder $\prec $ and an equivalence relation $\approx $ on the set $%
\mathcal{E}$ of all properties (for every $E$, $F$ $\in \mathcal{E}$, $%
E\prec F$ iff, for every $S\in \mathcal{S}$, $P_{S}(E)<P_{S}(F)$, and $%
E\approx F$ iff $E\prec F$ and $F\prec E$). Whenever $(\mathcal{E},\prec )$
is a lattice with an orthocomplementation $^{\bot }$ (i.e., a mapping of $%
\mathcal{E}$ into itself such that, for every $E\in \mathcal{E,}$ $E^{\bot
\bot }=E$, and for every $E$, $F$ $\in \mathcal{E}$, $E\prec F$ iff $F^{\bot
}\prec E^{\bot }$) and the mapping $P_{S}:E\in \mathcal{E}\rightarrow
P_{S}(E)\in \lbrack 0,1]$ satisfies some standard conditions (in particular,
the probability of the join of disjoint properties is the sum of the
probabilities of the properties), we say that $P_{S}$ \textit{is a
generalized probability measure} on $(\mathcal{E},\prec ,^{\bot })$ and call 
$P_{S}(E)$ the \textit{Q-probability} of $E$ given $S$. If $(\mathcal{E}%
,\prec ,^{\bot })$ is not Boolean, then $P_{S}$ is a non-classical
probability measure, hence we conclude that non-classical probabilities can
be obtained as derived notions in our classical probabilistic framework. It
is then easy to show that $P_{S}$, if it is non-classical, does not allow to
define in a canonical way a \textit{conditional Q-probability} of a property 
$E$ given $S$ and another property $F$. We are thus led to introduce a
non-standard definition of conditional Q-probability by considering
successive measurements of $E$ and $F$ (we avoid details here for the sake
of brevity).

Let us come to QM. If we assume that QM belongs to the class $\mathbb{T}%
^{\prime }$ of theories defined above, we obtain some important achievements.

First of all, the quantum probability of a property $E$ in a state $S$
(Born's rule) can be considered as the specific form that the Q-probability $%
P_{S}(E)$ takes in QM. Hence we conclude that the non-classical character of
quantum probability can be explained in classical terms by taking into
account $\mu $-contexts. In particular, quantum probability can be seen as a
derived notion in a classical probabilistic framework, hence it can be given
an epistemic rather than an ontic interpretation. To be precise, it can be
interpreted as an (indirect) measure of our lack of knowledge of the
physical situation at a microscopic level, both referring to preparation and
to measurement procedures.

Secondly, the reflexive, symmetric but not transitive binary relation of
compatibility on $E$ introduced in QM can be seen as the specific form that
the relation $k$ takes in QM, which provides a natural explanation of it in
a contextual framework.

Thirdly, the quantum notion of conditional probability can be considered as
the specific form that the conditional Q-probability takes in QM, so that
its non-standard features (in particular, the violation of the Bayes
theorem) is explained in terms of contextuality.

We add that our perspective is supported by some previous research in the
literature. Indeed, mean conditional probabilities and mean probability
measurements are conceptually similar to the \textit{universal averages} and
the \textit{universal measurements}, respectively, introduced by Aerts and
Sassoli de Bianchi (2014, 2017). Our recognition that two sources of
randomness underlie each measurement procedure also agrees with analogous
remarks of these authors.

To close, we note that we showed in (Garola, 2018) that CM can be included
in the class $\mathbb{T}^{\prime }$ as a special theory, in which truth
assignments on $\Psi (x)$ do not depend on the $\mu $-contexts and the
compatibility relation $k$ is trivial.

\section{The contextuality of quantum mechanics: a critical view}

We have shown in the previous sections that some bridges can be thrown
between the classical and the quantum views even if the standard
interpretation of QM is adopted. In particular, we have explicitly accepted
in Section 4 the standard view according to which QM is a contextual theory.
This view rests nowadays on the theorems quoted in Section 2.3, which are
maintained to prove that assumption O in Section 2.3 is untenable in QM,
hence to provide an irrefutable support to the thesis, going back to the
founders of the theory, that the properties of a quantum physical system are
only potential before a measurement (or non-objective, according to the
terminology adopted in this paper).

However, the conclusion above may be questioned. Indeed, we have shown in
some previous papers (see, e.g., Garola and Sozzo, 2010; Garola and Persano,
2014) that the aforesaid theorems rest on a hidden epistemological
assumption. If such assumption is accepted, then the standard view follows
at once, but if it is rejected, then new interpretations of QM become
possible, at least in principle, which avoid contextuality, hence the
non-objectivity of properties. We supply in this section a brief account of
this issue. The reader can find more complete treatment and bibliography in
the papers quoted above.

Let us consider Bell's (1966) and Kochen-Specker's (1967) theorems. These
theorems aim to show that (local) hidden variables that would determine the
values of all observables of a quantum physical system independently of any
measurement cannot exist (which amounts to say, from our present point of
view, that assumption O in Section 3.2 is untenable when dealing with
quantum physical systems), and many different proofs of them besides the
original ones have been provided. But an analysis of such proofs shows that
they rest, as stated above, on an epistemological assumption that is usually
left implicit. To make the point clear, let us consider a typical scheme of
proof of contextuality (see also Greenberger et al., 1990; Mermin, 1993).

(i) A theoretical law $\Lambda $ of QM is considered (e.g., the conservation
of total angular momentum in the case of a spin 1 particle).

(ii) Some laws $\Lambda _{1}$, $\Lambda _{2}$, \ldots\ are deduced from $%
\Lambda $, each of which contains only compatible observables and
establishes correlations among the possible values of the observables that
occur in it (via a \textit{Kochen-Specker condition} that we do not report
here for the sake of brevity). But the choice of $\Lambda _{1}$, $\Lambda
_{2}$, \ldots\ is done in such a way that there are observables in some laws
that are not compatible with observables that occur in other laws.

(iii) It is assumed (ad absurdum) that the values of the observables are
independent of the measurements that one can perform to check (the
predictions of) $\Lambda _{1}$, $\Lambda _{2}$, \ldots\ .

(iv) It is shown that, if $\Lambda _{1}$, $\Lambda _{2}$, \ldots\ are
suitably chosen, every assignment of values to all observables that occur in 
$\Lambda _{1}$, $\Lambda _{2}$, \ldots\ contradicts some of the correlations
among possible values established by these laws.

(v) The contradiction in (iv) follows from the assumption in (iii), which is
therefore untenable (hence the contextuality of QM).

Based on the conclusion in (v), the accepted doctrine states that, whenever
we perform a measurement intended to check one of the laws $\Lambda _{1}$, $%
\Lambda _{2}$, \ldots , say $\Lambda _{i}$, we determine a context that
actualizes some possible values of the (compatible) observables that occur
in $\Lambda _{i}$. These values are correlated in such a way that $\Lambda
_{i}$ is satisfied. But if we check another law, say $\Lambda _{j}$, in
which an observable $A$ occurs that also occurs in $\Lambda _{i}$, the
measurement context associated with $\Lambda _{j}$ may actualize a value of $%
A$ that is different from the value actualized by the measurement checking $%
\Lambda _{i}$.

To analyse critically the arguments above, let us refer to the received view
summarized in Section 2.1. According to this view, the laws $\Lambda _{1}$, $%
\Lambda _{2}$, \ldots\ can be considered as statements of the theoretical
language of QM, but $\Lambda _{1}$, $\Lambda _{2}$, \ldots\ correspond, via
the correspondence rules, to laws expressed by statements of the
observational language of QM, which we still denote by $\Lambda _{1}$, $%
\Lambda _{2}$, \ldots\ by abuse of language (hence $\Lambda _{1}$, $\Lambda
_{2}$, \ldots\ will be considered both as elements of the theoretical and of
the observational language). These laws are \textit{empirical laws}, in the
sense that each of them can be checked by means of measurements that
establish the values of the observables that occur in it. Every set of
measurements checking the law $\Lambda _{i}$ determines a (macroscopic) 
\textit{measurement context}. Thus we can associate with $\Lambda _{i}$ a
set of measurement contexts (there are generally many ways of checking $%
\Lambda _{i}$), and QM implies that $\Lambda _{i}$ is a true sentence in
each of these measurement contexts, so that the correlations among the
values of the observables established by $\Lambda _{i}$ (see (ii) above)
must be fulfilled by the outcomes of the measurements when $\Lambda _{i}$ is
checked. But the formalism of QM does not imply by itself that $\Lambda _{i}$
is true in those physical contexts in which it cannot be checked (e.g., in a
measurement context associated with a law $\Lambda _{j}\neq \Lambda _{i}$).
Stating that $\Lambda _{i}$ is true in every physical context is a
metalinguistic (epistemological) assumption on the laws of QM (\textit{%
assumption R} in the following) that can be added to the mathematical
formulation of QM but is not implied by it.

Now, reaching the conclusion in (v) requires just introducing assumption R
besides QM. Indeed, one needs to assume that $\Lambda _{1}$, $\Lambda _{2}$,
\ldots\ are true in every physical context to establish that one cannot
accept the contradiction in (iv) and must give up the assumption in (iii).
Equivalently, one needs to assume that the conjunction of $\Lambda _{1}$, $%
\Lambda _{2}$, \ldots\ must be always true, even if no measurement context
can be associated with it because non-compatible observables occur in it.
The introduction of assumption R in the reasoning above, however, usually
remains implicit, and most scholars are not aware of it, so that
contextuality is maintained to be a mere mathematical consequence of the
formalism of QM.

Of course, introducing assumption R is not wrong, and no criticism can be
done in this sense to the standard view. But it must be considered as an
(epistemological) choice, not a logical necessity. Moreover, it should be
noted that this choice is consistent with a classical realistic conception
of theoretical laws, even if its consequences are highly non-classical when
dealing with quantum physical systems, because it implies abandoning every
attempt at introducing a classical semantics in the language of QM.

One is thus led to inquire the alternative nonstandard choice of dispensing
with assumption R. One then concludes that, whenever this choice is made,
the conjunction of $\Lambda _{1}$, $\Lambda _{2}$, \ldots , if considered as
a sentence of the observational language, is not bound to be always true. It
joins indeed the empirical laws $\Lambda _{1}$, $\Lambda _{2}$, \ldots ,
each of which is true in its measurement contexts but may be false in
physical contexts in which it cannot be checked. All predictions of QM are
thus preserved, but the contradiction in (iv) does not imply that the
assumption in (iii) must be rejected. Hence the possibility of adopting a
classical semantics for the language of QM avoiding contextuality cannot be
excluded.

The above arguments suggest a generalization. Indeed, assumption R can be
seen as a special case of a general principle (which we called \textit{%
meta-theoretical classical principle}, or \textit{MCP}, in some previous
papers, see, e.g., Garola and Persano, 2014) that states that all physical
laws expressed as sentences of the observational language of a physical
theory must be true in every physical context. Then we call accepting MCP 
\textit{standard epistemological position} in the following.

Dispensing with R suggests instead a new general principle (which we called 
\textit{meta-theoretical generalized principle}, or \textit{MGP}, see again
Garola and Persano, 2014) that states that all physical laws expressed as
sentences of the observational language of a physical theory are bound to be
true in every physical context in which they can be checked, but can be true
as well as false in those physical contexts in which they cannot be checked.
Then, we call accepting MGP \textit{nonstandard epistemological position} in
the following. This position is consistent with interpreting the
mathematical apparatus of QM as a calculus whose role is producing laws of
the observational language by deduction and by using the correspondence
rules. Whenever one of these laws relates only compatible observables, some
measurement contexts exist in which the law is bound to be true (i.e., the
assignments of values to the observables that occur in it are bound to
satisfy the law itself) and in these contexts it can be checked. But if the
law relates incompatible observables, then it cannot be checked in any
physical context, and no measurement context exists in which the law is
bound to be true (i.e., the assignments of values to the observables that
occur in it are not bound to satisfy the law itself). Any realistic
conception of the theoretical laws of QM must thus be excluded, which makes
the foregoing nonstandard epistemological position more consistent with the
\textquotedblleft anti-metaphysical\textquotedblright\ attitude of QM than
the standard epistemological position.

Summing up, the standard view adopts a standard epistemological position,
hence accepts a widespread conception of the laws of physics, but is then
prevented from introducing a classical semantics for the language of QM
because of the non-objectivity of properties following from contextuality.
An alternative view adopting the nonstandard epistemological position
(which, even if nonstandard, is consistent with the received view, see
Section 2.1) does not meet with a similar prohibition, which opens the way
to alternative interpretations of QM that avoid non-objectivity of
properties, thus allowing the introduction of a classical semantics for the
language of QM. A new bridge is thus thrown between the classical and the
quantum views of the world.

Both the standard and the nonstandard epistemological positions meet,
however, some problems. The former implies, in particular, contextuality at
a distance, or nonlocality, which is counterintuitive. The latter may avoid
this kind of contextuality but implies a kind of contextuality of empirical
physical laws that can be seen as a consequence of the limits of our
theoretical knowledge but is somewhat uneasy. However, the nonstandard
epistemological position has the merit of leaving open the possibility of
finding a more general theoretical framework, embedding QM, in which
properties are objective. In this framework QM would be regarded as an
incomplete theory, recovering a perspective going back to Einstein (1935).

To close, we remind that we have tried in several papers to take some steps
towards the construction of the theoretical framework envisaged above (see,
e.g., Garola, 2015; Garola et al., 2016). Our proposal (which we called 
\textit{ESR model}) assumes that any property that a microscopic physical
system may possess is either possessed or not possessed by the system (here
called \textit{physica}l \textit{object}, as in the previous sections)
independently of any measurement, and tries to explain the laws of QM in
terms of more general laws. The basic idea consists in assigning a
probability that a physical object may remain undetected when a measurement
on it is performed because of the set of properties possessed by the
physical object itself rather than because of a lack of efficiency of the
measurements. The quantum probability that a physical object in a state $S$
turns out to possess the property $E$ when a measurement of $E$ is performed
on it would then refer to the set of physical objects that can be detected
in this kind of measurement rather than to the set of physical objects that
are prepared in the state $S$. Thus, our ESR model modifies the standard
interpretation of quantum probabilities but preserves the mathematical
apparatus of QM, which is embedded into a more general mathematical
framework. Moreover, the ESR model predicts that there are upper limits
(depending on the measurements that are considered) to the percentage of
objects that can be detected, which makes it falsifiable.\medskip 

\textbf{Acknowledgement}. The author is greatly indebted with Dr. Antonio
Negro for carefully reading the manuscript and providing many useful remarks
and suggestions.

\begin{center}
\textbf{BIBLIOGRAPHY}
\end{center}

Aerts, D. (1999). Foundations of quantum physics: a general realistic and
operational approach. \textit{Int. J. Theor. Phys.} \textbf{38}, 289-358.

Aerts, D. and Sassoli de Bianchi, M. (2014). The extended Bloch
representation of quantum mechanics and the hidden-measurement solution of
the measurement problem. \textit{Ann. Phys}. \textbf{351}, 975-1025.

Aerts, D. and Sassoli de Bianchi, M. (2017). \textit{Universal Measurements.
How to Free Three Birds in One Move.} World Scientific, Singapore.

Bell, J.S. (1964). On the Einstein-Podolski-Rosen Paradox. \textit{Physics} 
\textbf{1}, 195--200.

Bell, J.S. (1966). On the Problem of Hidden Variables in Quantum Mechanics. 
\textit{Rev. Mod. Phys.} \textbf{38}, 447--452.

Birkhoff, G. and von Neumann, J. (1936). The Logic of Quantum Mechanics. 
\textit{Ann. Math.} \textbf{37}, 823--843.

Bohr, N. (1958). \textit{Atomic Physics and Human Knowledge}. John Wiley and
Sons, London.

Braithwaite, R.B. (1953). \textit{Scientific Explanation}. Cambridge
University Press, Cambridge.

Busch, P., Lahti, P.J. and Mittelstaedt, P. (1996). \textit{The Quantum
Theory of Measurement}. Springer, Berlin.

Carnap, R. (1966). \textit{Philosophical Foundations of Physics}. Basic
Books Inc., New York.

Dalla Chiara, M. L., Giuntini, R. and Greechie, R. (2004). \textit{Reasoning
in Quantum Theory}. Kluwer, Dordrecht.

Dalla Pozza, C. and Garola, C. (1995). A pragmatic interpretation of
intuitionistic propositional logic. \textit{Erkenntnis} \textbf{43}, 81-109.

Einstein, A., Podolski, B. and Rosen, N. (1935). Can quantum mechanical
description of physical reality be considered complete? \textit{Phys. Rev.} 
\textbf{47}, 777-780.

Feyerabend, F. (1975). \textit{Against Method: Outline of an Anarchist
Theory of Knowledge}. New Left Books, London.

Feynmann, R.P. (1964). The Messenger Lecture Series at Cornell, Lecture 6.
In \textit{The Character of Physical Laws}. The MIT Press, 1967/1917.

Frege, G. (1893). \textit{Grundgesetze der Arithmetik I}, Pohle, Jena.

Garola, C. (1992). Truth versus testability in quantum logic. \textit{%
Erkenntnis} \textbf{37}, 197-222.

Garola, C. (2008). Physical propositions and quantum languages. \textit{Int.
J. Theor. Phys.} \textbf{47}, 90-103.

Garola, C. (2015). A survey of the ESR model for an objective interpretation
of quantum mechanics. \textit{Int. J. Theor. Phys.} \textbf{54}, 4410-4422.

Garola, C. (2017). Interpreting quantum logic as a pragmatic structure, 
\textit{Int. J. Theor. Phys.} \textbf{56}, 3770-3782.

Garola, C. (2018). An epistemic interpretation of quantum probability via
contextuality. \textit{Found. Sci.,} DOI: 10.1007/s10699-018-9560-4.

Garola, C. and Persano, M. (2014). Embedding quantum mechanics into a
broader noncontextual theory. \textit{Found. Sci.} \textbf{19}, 217-239.

Garola, C. and Sozzo, S. (2010). Realistic aspects in the standard
interpretation of quantum mechanics. \textit{Humana.ment. J. Phil. Stud.} 
\textbf{13}, 81-101.

Garola, C. and Sozzo, S. (2013). Recovering quantum logic within an extended
classical framework. \textit{Erkenntnis} \textbf{78}, 399-314.

Garola, C., Sozzo, S. and Wu, J. (2016). Outline of a generalization and a
reinterpretation of quantum mechanics recovering objectivity. \textit{Int.
J. Theor. Phys.} \textbf{55}, 2500-2528.

Greenberger, D.M., Horne, M.A., Shimony, A. and Zeilinger, A. (1990). Bell's
theorem without inequalities. \textit{Am. J. Phys.} \textbf{58}, 1131-1143.

Haack, S. (1974). \textit{Deviant Logic}. Cambridge University Press,
Cambridge.

Haack, S. (1978).\textit{\ Philosophy of Logic}. Cambridge University Press,
Cambridge.

Heisenberg, W. (1958). Physics and Philosophy: the Revolution of Modern
Science. Harper, New York.

Hempel, C.C. (1965). \textit{Aspects of Scientific Explanation}. Free Press,
New York.

Kochen, S. and Specker, E. P. (1967). The Problem of Hidden Variables in
Quantum Mechanics. \textit{J. Math. Mech.} \textbf{17}, 59--87.

Kuhn, T.S. (1962). \textit{The Structure of Scientific Revolution}. Chicago
University Press, Chicago.

Mermin, N.D. (1993). Hidden variables and the two theorems of John Bell. 
\textit{Rev. Mod. Phys.} \textbf{65}, 803-815.

Nagel, E. (1961). \textit{The Structure of Science}. Harcourt, Brace \&
World, New York.

Piron, C. (1976). \textit{Foundations of Quantum Physics}. Benjamin, Reading
(MA).

R\'{e}dei, N. (1998). \textit{Quantum Logic in Algebraic Approach.} Kluwer,
Dordrecht.

Schlosshauer, M., Kofler, J., Zeilinger, A. (2013). A snapshot of
foundational attitudes toward quantum mechanics. \textit{Stud. Hist. Phil.
Mod. Phys. }\textbf{44} (3), 222-230.

Tarski, A. (1944). The semantic conception of truth and the foundations of
semantics. \textit{Philosophy and Phenomenological Research} \textbf{4},
341-375.

Tarski, A. (1956). The concept of truth in formalized languages. In J.M.
Woodger (Ed.), \textit{Logic, Semantics, Metamathematics} (pp. 152-268).
Oxford University Press, Oxford.

\end{document}